\documentclass[12pt,a4paper]{article}

\usepackage{ifthen} 
\newboolean{pdflatex}
\setboolean{pdflatex}{true} 

\newboolean{articletitles}
\setboolean{articletitles}{true} 

\newboolean{uprightparticles}
\setboolean{uprightparticles}{false} 

\newboolean{inbibliography}
\setboolean{inbibliography}{false} 

\usepackage[top=1in, bottom=1.25in, left=1in, right=1in]{geometry}

\usepackage{microtype}
\usepackage{lineno}  
\usepackage{xspace} 
\usepackage{caption} 

\usepackage{graphicx}  
\usepackage{color}
\usepackage{colortbl}
\graphicspath{{./figs/}} 

\usepackage{amsmath} 
\usepackage{amssymb}
\usepackage{amsfonts}
\usepackage{upgreek} 

\newcommand*\patchAmsMathEnvironmentForLineno[1]{%
\expandafter\let\csname old#1\expandafter\endcsname\csname #1\endcsname
\expandafter\let\csname oldend#1\expandafter\endcsname\csname
end#1\endcsname
 \renewenvironment{#1}%
   {\linenomath\csname old#1\endcsname}%
   {\csname oldend#1\endcsname\endlinenomath}%
}
\newcommand*\patchBothAmsMathEnvironmentsForLineno[1]{%
  \patchAmsMathEnvironmentForLineno{#1}%
  \patchAmsMathEnvironmentForLineno{#1*}%
}
\AtBeginDocument{%
\patchBothAmsMathEnvironmentsForLineno{equation}%
\patchBothAmsMathEnvironmentsForLineno{align}%
\patchBothAmsMathEnvironmentsForLineno{flalign}%
\patchBothAmsMathEnvironmentsForLineno{alignat}%
\patchBothAmsMathEnvironmentsForLineno{gather}%
\patchBothAmsMathEnvironmentsForLineno{multline}%
\patchBothAmsMathEnvironmentsForLineno{eqnarray}%
}

\usepackage{hyperref}    
\usepackage[all]{hypcap} 

\usepackage{xspace} 
\usepackage{upgreek}

\def\lhcb {\mbox{LHCb}\xspace}

\def\MagUp {\mbox{\em Mag\kern -0.05em Up}\xspace}

\ifthenelse{\boolean{uprightparticles}}%
{

 \def\Pmu         {\ensuremath{\upmu}\xspace}

 \def\Ppi         {\ensuremath{\uppi}\xspace}

 \def\Pphi        {\ensuremath{\upphi}\xspace}

 \def\Ppsi        {\ensuremath{\uppsi}\xspace}

 \def\PDelta      {\ensuremath{\Delta}\xspace}                 
 \def\PXi      {\ensuremath{\Xi}\xspace}                 
 \def\PLambda      {\ensuremath{\Lambda}\xspace}                 
 \def\PSigma      {\ensuremath{\Sigma}\xspace}                 
 \def\POmega      {\ensuremath{\Omega}\xspace}                 
 \def\PUpsilon      {\ensuremath{\Upsilon}\xspace}                 
 
 \def\PB      {\ensuremath{\mathrm{B}}\xspace}                 
                  
 \def\PD      {\ensuremath{\mathrm{D}}\xspace}

 \def\PJ      {\ensuremath{\mathrm{J}}\xspace}                 
 \def\PK      {\ensuremath{\mathrm{K}}\xspace}

 \def\Pb      {\ensuremath{\mathrm{b}}\xspace}                 
 \def\Pc      {\ensuremath{\mathrm{c}}\xspace}

 \def\Pi      {\ensuremath{\mathrm{i}}\xspace}

}
{

 \def\Pmu         {\ensuremath{\mu}\xspace}

 \def\Ppi         {\ensuremath{\pi}\xspace}

 \def\Pphi        {\ensuremath{\phi}\xspace}

 \def\Ppsi        {\ensuremath{\psi}\xspace}                 
                  
 \mathchardef\PDelta="7101
 \mathchardef\PXi="7104
 \mathchardef\PLambda="7103
 \mathchardef\PSigma="7106
 \mathchardef\POmega="710A
 \mathchardef\PUpsilon="7107
                  
 \def\PB      {\ensuremath{B}\xspace}                 
                  
 \def\PD      {\ensuremath{D}\xspace}

 \def\PJ      {\ensuremath{J}\xspace}                 
 \def\PK      {\ensuremath{K}\xspace}

 \def\Pb      {\ensuremath{b}\xspace}                 
 \def\Pc      {\ensuremath{c}\xspace}

 \def\Pi      {\ensuremath{i}\xspace}

}

\makeatletter
\ifcase \@ptsize \relax
  \newcommand{\miniscule}{\@setfontsize\miniscule{4}{5}}
\or
  \newcommand{\miniscule}{\@setfontsize\miniscule{5}{6}}
\or
  \newcommand{\miniscule}{\@setfontsize\miniscule{5}{6}}
\fi
\makeatother

\DeclareRobustCommand{\optbar}[1]{\shortstack{{\miniscule (\rule[.5ex]{1.25em}{.18mm})}
  \\ [-.7ex] $#1$}}

\def\mumu       {{\ensuremath{\Pmu^+\Pmu^-}}\xspace}

\def\cquark    {{\ensuremath{\Pc}}\xspace}

\def\bquark    {{\ensuremath{\Pb}}\xspace}

\def\pion   {{\ensuremath{\Ppi}}\xspace}

\def\pip    {{\ensuremath{\pion^+}}\xspace}
\def\pim    {{\ensuremath{\pion^-}}\xspace}

\def\kaon    {{\ensuremath{\PK}}\xspace}
  \def\Kbar    {{\kern 0.2em\overline{\kern -0.2em \PK}{}}\xspace}

\def\KorKbar    {\kern 0.18em\optbar{\kern -0.18em K}{}\xspace}

\def\Kp      {{\ensuremath{\kaon^+}}\xspace}

\def\KS      {{\ensuremath{\kaon^0_{\mathrm{ \scriptscriptstyle S}}}}\xspace}

\def\Kstarz  {{\ensuremath{\kaon^{*0}}}\xspace}

\newcommand{\phiz}{\ensuremath{\Pphi}\xspace}

  \def\Dbar    {{\kern 0.2em\overline{\kern -0.2em \PD}{}}\xspace}
\def\D       {{\ensuremath{\PD}}\xspace}

\def\DorDbar    {\kern 0.18em\optbar{\kern -0.18em D}{}\xspace}
\def\Dz      {{\ensuremath{\D^0}}\xspace}

\def\B       {{\ensuremath{\PB}}\xspace}
\def\Bbar    {{\ensuremath{\kern 0.18em\overline{\kern -0.18em \PB}{}}}\xspace}

\def\BorBbar    {\kern 0.18em\optbar{\kern -0.18em B}{}\xspace}
\def\Bz      {{\ensuremath{\B^0}}\xspace}

\def\Bu      {{\ensuremath{\B^+}}\xspace}

\def\Bp      {{\ensuremath{\Bu}}\xspace}

\def\Bdb     {{\ensuremath{\Bbar{}^0}}\xspace}

\def\jpsi     {{\ensuremath{{\PJ\mskip -3mu/\mskip -2mu\Ppsi\mskip 2mu}}}\xspace}
\def\psitwos  {{\ensuremath{\Ppsi{(2S)}}}\xspace}
\def\psiprpr  {{\ensuremath{\Ppsi(3770)}}\xspace}

  \def\Y#1S{\ensuremath{\PUpsilon{(#1S)}}\xspace}

\def\Lz          {{\ensuremath{\PLambda}}\xspace}
\def\Lbar        {{\ensuremath{\kern 0.1em\overline{\kern -0.1em\PLambda}}}\xspace}
\def\LorLbar    {\kern 0.18em\optbar{\kern -0.18em \PLambda}{}\xspace}

\def\BF         {{\ensuremath{\mathcal{B}}}\xspace}

\def\to                 {\ensuremath{\rightarrow}\xspace}

\def\AT#1     {\ensuremath{A_{\mathrm{T}}^{#1}}\xspace}           

\def\C#1      {\ensuremath{\mathcal{C}_{#1}}\xspace}                       
\def\Cp#1     {\ensuremath{\mathcal{C}_{#1}^{'}}\xspace}                    
\def\Ceff#1   {\ensuremath{\mathcal{C}_{#1}^{\mathrm{(eff)}}}\xspace}        
\def\Cpeff#1  {\ensuremath{\mathcal{C}_{#1}^{'\mathrm{(eff)}}}\xspace}       
\def\Ope#1    {\ensuremath{\mathcal{O}_{#1}}\xspace}                       
\def\Opep#1   {\ensuremath{\mathcal{O}_{#1}^{'}}\xspace}                    

\newcommand{\tev}{\ifthenelse{\boolean{inbibliography}}{\ensuremath{~T\kern -0.05em eV}\xspace}{\ensuremath{\mathrm{\,Te\kern -0.1em V}}}\xspace}
\newcommand{\gev}{\ensuremath{\mathrm{\,Ge\kern -0.1em V}}\xspace}
\newcommand{\mev}{\ensuremath{\mathrm{\,Me\kern -0.1em V}}\xspace}
\newcommand{\kev}{\ensuremath{\mathrm{\,ke\kern -0.1em V}}\xspace}
\newcommand{\ev}{\ensuremath{\mathrm{\,e\kern -0.1em V}}\xspace}
\newcommand{\gevc}{\ensuremath{{\mathrm{\,Ge\kern -0.1em V\!/}c}}\xspace}
\newcommand{\mevc}{\ensuremath{{\mathrm{\,Me\kern -0.1em V\!/}c}}\xspace}
\newcommand{\gevcc}{\ensuremath{{\mathrm{\,Ge\kern -0.1em V\!/}c^2}}\xspace}
\newcommand{\gevgevcccc}{\ensuremath{{\mathrm{\,Ge\kern -0.1em V^2\!/}c^4}}\xspace}
\newcommand{\mevcc}{\ensuremath{{\mathrm{\,Me\kern -0.1em V\!/}c^2}}\xspace}

\def\cm   {\ensuremath{\mathrm{ \,cm}}\xspace}

\def\invfb   {\ensuremath{\mbox{\,fb}^{-1}}\xspace}

\def\ps   {\ensuremath{{\mathrm{ \,ps}}}\xspace}

\def\gsim{{~\raise.15em\hbox{$>$}\kern-.85em
          \lower.35em\hbox{$\sim$}~}\xspace}
\def\lsim{{~\raise.15em\hbox{$<$}\kern-.85em
          \lower.35em\hbox{$\sim$}~}\xspace}

\def\sqs   {\ensuremath{\protect\sqrt{s}}\xspace}

\def\pt         {\mbox{$p_{\mathrm{ T}}$}\xspace}

\def\tell1  {TELL1\xspace}
\def\ukl1   {UKL1\xspace}

\newcommand{\ie}{\mbox{\itshape i.e.}\xspace}

\usepackage{cite} 
\usepackage{mciteplus}

\usepackage{longtable} 

\begin{document}

\renewcommand{\thefootnote}{\fnsymbol{footnote}}
\setcounter{footnote}{1}

\begin{titlepage}
\pagenumbering{roman}

\vspace*{-1.5cm}
\centerline{\large EUROPEAN ORGANIZATION FOR NUCLEAR RESEARCH (CERN)}
\vspace*{1.5cm}
\noindent
\begin{tabular*}{\linewidth}{lc@{\extracolsep{\fill}}r@{\extracolsep{0pt}}}
\ifthenelse{\boolean{pdflatex}}
{\vspace*{-2.7cm}\mbox{\!\!\!\includegraphics[width=.14\textwidth]{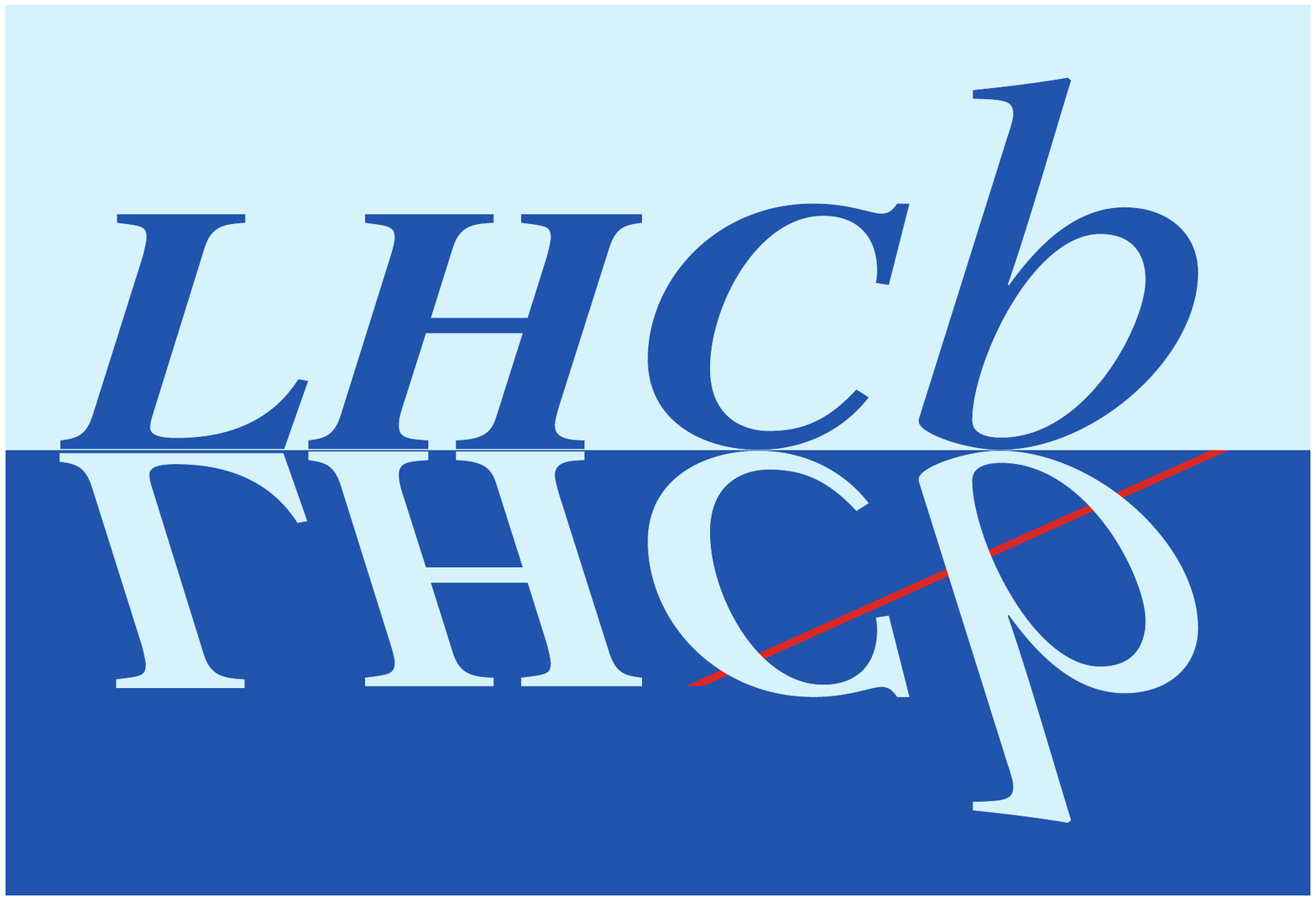}} & &}%
{\vspace*{-1.2cm}\mbox{\!\!\!\includegraphics[width=.12\textwidth]{lhcb-logo.eps}} & &}%
\\
 & & CERN-EP-2016-302 \\  
 & & LHCb-PAPER-2016-052 \\  
 & & \today \\ 
 & & \\
\end{tabular*}

\vspace*{3.8cm}

{\normalfont\bfseries\boldmath\huge
\begin{center}
  Search for long-lived scalar particles in $B^+ \to K^+ \chi (\mumu)$ decays
\end{center}
}

\vspace*{1.9cm}

\begin{center}
The LHCb collaboration\footnote{Authors are listed at the end of this paper.}
\end{center}

\vspace{\fill}

\begin{abstract}
  \noindent
A search for a long-lived scalar particle $\chi$ is performed, looking for the decay ${\PB^+ \to \PK^+ \chi}$ with ${\chi \to \mumu}$ in $pp$ collision data corresponding to an integrated luminosity of $3\,\invfb$, collected by the LHCb experiment at centre-of-mass energies of $\sqrt{s}=7$ and 8$\,$TeV. This new scalar particle, predicted by Hidden Sector models, is assumed to have a narrow width. 
The signal would manifest itself as an excess in the dimuon invariant mass distribution over the Standard Model background.
No significant excess is observed in the accessible ranges of mass ${250 < m(\chi) < 4700\,\mevcc}$ and lifetime ${0.1 < \tau(\chi) < 1000\,\ps}$.
Upper limits on the branching fraction $\BF(\Bp \to \Kp \chi (\mumu))$ at 95\% confidence level are set as a function of $m(\chi)$ and $\tau(\chi)$, varying between $2\times10^{-10}$ and $10^{-7}$. These are the most stringent limits to date. 
The limits are interpreted in the context of a model with a light inflaton particle. 

\end{abstract}

\vspace*{2.0cm}

\begin{center}
  Published in Phys.~Rev.~D
\end{center}

\vspace{\fill}

{\footnotesize 
\centerline{\copyright~CERN on behalf of the \lhcb collaboration, licence \href{http://creativecommons.org/licenses/by/4.0/}{CC-BY-4.0}.}}
\vspace*{2mm}

\end{titlepage}


\newpage
\setcounter{page}{2}
\mbox{~}
%
%
%
%

\cleardoublepage

\renewcommand{\thefootnote}{\arabic{footnote}}
\setcounter{footnote}{0}



\pagestyle{plain} 
\setcounter{page}{1}
\pagenumbering{arabic}


%

In recent years, models with a Hidden Sector of particles~\cite{Essig:2013lka,Batell:2009jf} have gathered considerable attention, primarily motivated by an absence of direct dark matter identification. 
This class of theories postulates the existence of new particles that interact very weakly with the particles of the Standard Model (SM). 
In this scenario, dark-sector particles would be gauge-singlet states with respect to the SM gauge group, 
and only be able to communicate with SM particles via weakly interacting mediators through one of four mechanisms: the vector, axion, Higgs, and neutrino portals. 

\begin{figure}[h]
  \begin{center}
    \includegraphics[width=0.49\linewidth]{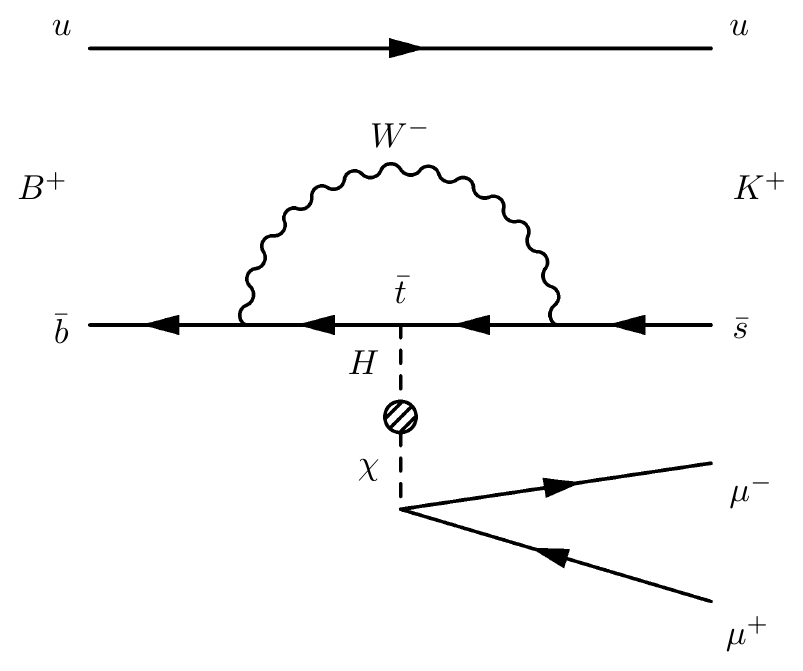}
    \vspace*{-0.7cm}
  \end{center}
  \caption{Feynman diagram of the decay $\Bp \to \Kp \chi (\mumu )$, where the $\chi$ interacts by mixing with the Higgs and then decays to a pair of muons.
    }
  \label{fig:Feynman_diag}
\end{figure}

In the Higgs portal scenario, the new scalar particle, $\chi$,  can mix with the SM Higgs boson. 
An example of such a model is described in Refs.~\cite{inflatonGuide,Bezrukov:2013fca}. 
In this theory, the Higgs portal is mediated by a light particle, namely the inflaton, associated to the field that generates the inflation of the early Universe.
These models also help to solve the hierarchy problem and can explain the baryon asymmetry in the Universe~\cite{Hertzberg:2013mba,Hertzberg:2013jba}. 
The inflaton mass and lifetime are weakly constrained; in particular, the mass can be below the $\PB$ meson mass, and the decay of $\Bp \to \Kp \chi$, with $\chi \to \mumu$, is a candidate process in which to look for such phenomena at LHCb. 
As illustrated in Figure~\ref{fig:Feynman_diag}, in this scenario the inflaton couples via the Higgs boson to the top quark that at loop level mediates the $\Bp$ to $\Kp$ transition.

%
%
Current limits on the process have been set by the CHARM experiment~\cite{CHARM} and, looking for $\Bz \to \Kstarz \chi (\mumu)$ decays, the \lhcb experiment~\cite{LHCB-PAPER-2015-036}. 
This Letter presents the search for a hypothetical new scalar particle through the decay $\Bp \to \Kp \chi (\mumu )$ in the ranges of mass $250 < m(\chi) < 4700\,\mevcc$ and lifetime $0.1 < \tau(\chi) < 1000\,\ps$. The inclusion of charge-conjugate decays is implied throughout this Letter.
The data sample used in this analysis corresponds to integrated luminosities of 1 and $2\,\invfb$ collected by the \lhcb detector in $pp$ collisions at centre-of-mass energies of $\sqs = 7$ and $8\,\tev$, respectively.

The \lhcb detector~\cite{Alves:2008zz,LHCb-DP-2014-002} is a single-arm forward
spectrometer covering the \mbox{pseudorapidity} range $2<\eta <5$,
designed for the study of particles containing \bquark or \cquark
quarks. 
The detector has a silicon-strip vertex detector as the first component of a 
high-precision charged-particle tracking system for measuring momenta;
two ring-imaging Cherenkov detectors for distinguishing charged hadrons;
a calorimeter system for identifying photons, electrons, and hadrons;
and a system for identifying muons.
The online event selection is performed by a trigger 
consisting of a hardware stage, based on information from the calorimeter and muon
systems, followed by a software stage, which applies a full event
reconstruction.

The hardware trigger selects events containing at least one muon with large transverse momentum (\pt)~\cite{Albrecht:2013fba}. 
The software trigger requires a two- or three-track secondary vertex with a significant displacement from the primary $pp$-interaction vertices (PVs). 
Finally, the reconstructed \Bp decay vertex is required to be significantly displaced from all PVs.
Only tracks with segments reconstructed in the vertex detector are considered by the trigger algorithms, \ie the $\chi$ boson is required to decay within a distance of about $60\,\cm$ from the PV.


In the simulation, $pp$ collisions are generated following Refs.~\cite{Sjostrand:2006za, *Sjostrand:2007gs, LHCb-PROC-2010-056, Lange:2001uf}, and the
interaction of the generated particles with the detector, and its response,
are implemented as in Refs. ~\cite{Allison:2006ve, *Agostinelli:2002hh, LHCb-PROC-2011-006}.
Signal decays are generated using a phase-space model. 
Simulations were performed for discrete values of $m(\chi)$ and $\tau(\chi)$,
and the resulting efficiencies interpolated to span the range covered
in the analysis.




In the search presented in this Letter, the dimuon mass distribution, $m(\mumu)$, is scanned in steps of half of the dimuon mass resolution, looking for a significant excess over the expected background yield. 
%
%
Three regions of dimuon decay time, $t(\mumu)$, are defined in the search: 
a prompt region, $|t(\mumu)| < 1\,\ps$,  
an intermediate region, $1 < t(\mumu) < 10\,\ps$, 
and an extremely displaced region, $t(\mumu) > 10\,\ps$.
The ranges of the three regions are optimized to provide the tightest upper limits.
The strategy, based on the approach of Ref.~\cite{Williams:2015xfa}, is similar to that used in Ref.~\cite{LHCB-PAPER-2015-036}.


The branching fraction of the $\Bp \to \Kp \chi (\mumu)$ decay is normalized to the well known $\Bp \to \Kp \jpsi (\mumu)$ decay, 
which has a branching fraction of ${\BF(\Bp \to \Kp \jpsi (\mumu)) = (6.12 \pm 0.19) \times 10^{-5}}$~\cite{PDG2016}.
In order to avoid experimental biases, a blind analysis is performed, in which the analysis is optimized without examining the $\Bp \to \Kp \chi (\mumu)$ candidates that have an invariant mass close to the known \Bp mass~\cite{PDG2016}. 

In a first step, a loose candidate selection with the following requirements is applied:
the \Bp decay vertex is significantly separated from the PV; the \Bp candidate impact parameter (IP) is small, and the IP of the charged kaon and muons are large; the angle between the \Bp momentum vector and the vector between the PV and the \Bp decay vertex is small; and the kaon and the muons must each satisfy loose particle identification requirements.

To further reduce the level of combinatorial background, $\Bp$ candidates satisfying these requirements are filtered by a multivariate selection using a boosted decision tree~\cite{Breiman,AdaBoost}.
The inputs to the algorithm include the \pt and the decay time of the \Bp candidate, topological variables like the quality of the \Bp and $\chi$ vertices, their separation and impact parameters of the three tracks of the decay and two isolation criteria~\cite{LHCB-PAPER-2015-025, LHCB-PAPER-2013-046}. These variables show a good agreement between data and simulation.
Data from the high-mass sideband, $5450 < m(\Kp \mumu) < 5800\,\mevcc$, are employed as the background training sample, using the K-folding technique~\cite{k-fold} with K=11 folds.
A small dependence on the mass and lifetime of the signal training sample is observed in the performance of the multivariate selection.
The simulated signal sample generated with $m(\chi) = 2500\,\mevcc$ and $\tau(\chi) = 1\,\ps$ provides the best overall sensitivity and is used for training the boosted decision tree.
The candidate selection based on the classifier of the multivariate analysis is optimized separately in each decay-time region of the search, which results in a signal efficiency between 65\% and 97\%, depending on $m(\chi)$ and $\tau(\chi)$, and a combinatorial background rejection rate of 98\% in the first and second decay-time region, 
and 90\% in the third decay-time region of the search.

Besides the combinatorial background, the main background consists of SM ${\Bp \to \Kp \mumu}$ decays, 
which have the same final state as the signal. 
The decay-time resolution is studied with fully simulated events and is found to be
between 0.1 and $0.2\,\ps$, depending on the dimuon mass. 
Since the value of the decay-time resolution is much smaller than the boundary of the first decay-time region, 
$\Bp \to \Kp \mumu$ decays only affect the prompt decay-time region of the analysis.

Peaking backgrounds that survive the multivariate selection are vetoed explicitly.
Narrow SM dimuon resonances are vetoed by excluding the regions near the \phiz, \jpsi, \psitwos, \psiprpr and $\psi(4160)$ resonances.
Since the contributions of the \phiz and $\psi(4160)$ are negligible in the two displaced decay-time regions, they are only vetoed in the first decay-time region.
Candidates are also rejected if compatible with $\KS \to \pip \pim$ and $\Lz \to p \pim$ decays, 
when pion or proton masses are assigned to the final-state particles that have been identified as muon candidates.
Background from $\Dz \to \Kp \pim$ decays is rejected by tighter muon identification criteria when, after assignment of the
kaon and the pion mass to the final-state particles, the invariant mass is close to the \Dz mass.
Similarly, additional particle identification criteria are required to reject $\Bp \to \Kp \jpsi (\mumu)$ decays where the kaon is misidentified as a muon and the same-sign muon as a kaon.
All other particle misidentification backgrounds have been found to be negligible.

\begin{figure}[tb]
  \begin{center}
    \includegraphics[width=0.58\linewidth]{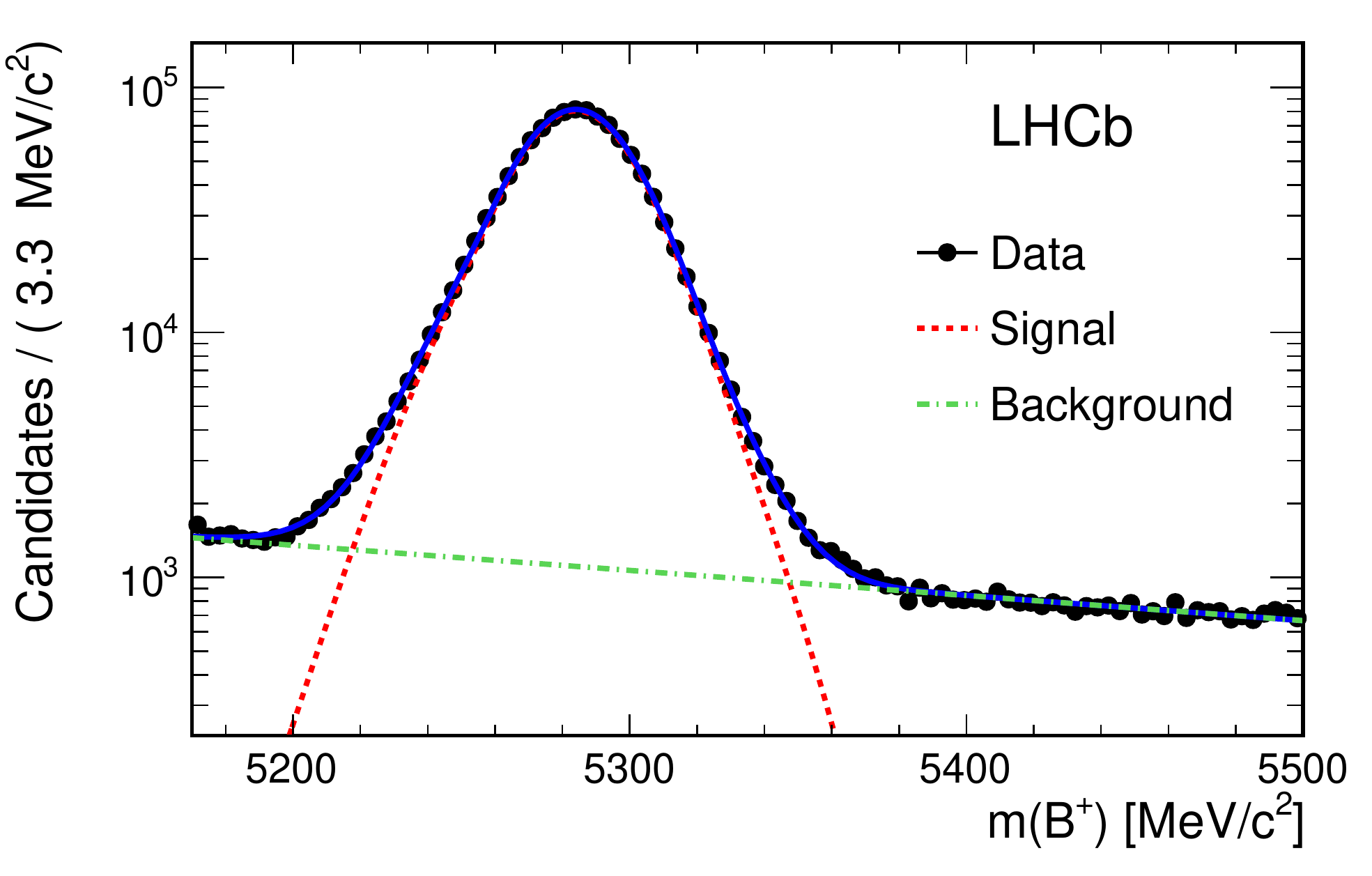}
    \vspace*{-0.7cm}
  \end{center}
  \caption{Invariant mass distribution for the normalization channel $\Bp \to \Kp \jpsi (\mumu)$ with the results of the fit overlaid. The signal is parametrised with a double-sided Crystal Ball (dotted red line)~\cite{LHCB-PAPER-2011-013} and the background with an exponential function (dashed green line).}
  \label{fig:normalization}
\end{figure}

Figure~\ref{fig:normalization} shows the invariant mass distribution for the $\Bp \to \jpsi(\mumu) \Kp$ normalization channel.
An extended unbinned maximum likelihood fit is performed to estimate the normalization yield.
The signal is parametrised with a double-sided Crystal Ball function~\cite{LHCB-PAPER-2011-013}, the background with an exponential function, with all parameters free to vary.
The fit yields $N(\Bp \to \jpsi(\mumu) \Kp) = (1142.0 \pm 0.4) \times 10^3$ signal events, where the uncertainty is only statistical.
 

\begin{figure}[tb]
  \begin{center}
    \includegraphics[width=0.72\linewidth]{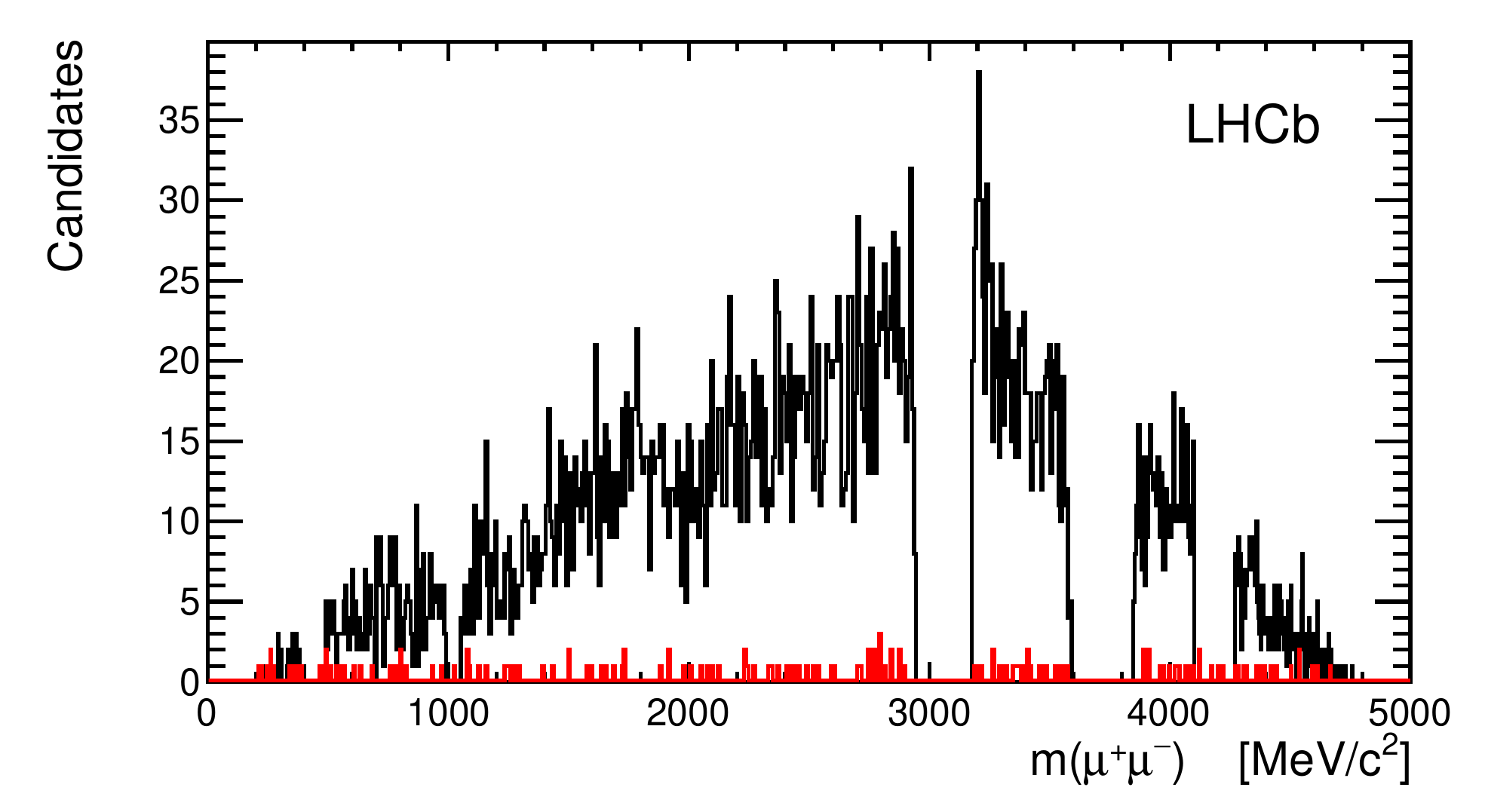}
    \vspace*{-0.7cm}
  \end{center}
  \caption{Distribution of the $m(\mumu)$ in the first (black) and second (red) decay-time region of the search. The binning scheme reflects the dimuon mass scanning procedure and the bin width corresponds to the mass-dependent dimuon resolution. Empty regions correspond to the \KS, \jpsi, \psitwos and \psiprpr vetoes (for both distributions) and  to the \phiz and $\psi(4160)$ vetoes (only for the prompt decay-time region).} 
  \label{fig:dimuon_mass}
\end{figure}

Figure~\ref{fig:dimuon_mass} shows the dimuon mass distributions in the first and second decay-time regions for candidates with invariant mass within $50\,\mevcc$ from the known \Bp mass~\cite{PDG2016}. 
The \Bp mass is constrained~\cite{Hulsbergen:2005pu} to its known value~\cite{PDG2016} to improve the dimuon mass resolution to be between 3 and $9\,\mevcc$.
No candidates populate the third decay-time region.
For each value of $m(\chi)$ and $\tau(\chi)$, the background plus signal and the background-only hypotheses
are compared using the CLs method~\cite{CLs1,CLs2}, where the information from the three decay-time regions is combined.
For each decay-time region, the expected background is obtained by linear interpolation of the dimuon mass sidebands~\cite{Williams:2015xfa}, 
while the shape of the mass distribution for the signal is taken from simulation.
No significant signal excess compared to the background-only hypothesis is found.


Several sources of systematic uncertainties are taken into account in setting upper limits.
Deviations from the linear approximation assumed in the background modelling are studied by means of pseudoexperiments.
This uncertainty is estimated to be 8\% of the statistical uncertainty and is assigned as a systematic uncertainty on the expected background yield.
Uncertainties affecting the expected signal yield strongly depend on $m(\chi)$ and $\tau(\chi)$. 
Both the signal efficiency and the mass resolution are computed from the simulation and validated on data using the control modes $\Bp \to \KS \jpsi(\mumu)$ and $\Bp \to \Kp \jpsi(\mumu)$. The decays $\Bp \to \Kp \psitwos (\mumu)$ and $\Bp \to \Kp \phiz (\mumu)$ are studied in data and the observed yields are found to be compatible with the expected SM branching fractions of $(4.9 \pm 0.6) \times 10^{-6}$ and $(2.5 \pm 0.3) \times 10^{-9}$, respectively~\cite{PDG2016}.
The main sources of uncertainty can be attributed to the uncertainty in the branching fraction of the normalisation channel, contributing 3\%, to the limited sizes of the simulated signal samples, which gives an uncertainty between 2 and 6\% of the expected signal yield, to the signal mass resolution, which contributes between 1.5 and 2\%, and to the lifetime extrapolation from the simulation.
The last, based on reweighting the events generated with $\tau(\chi) = 100\,\ps$ to match the lifetime distribution in the studied range, results in uncertainties between 0 and 20\% for $\tau(\chi) > 100\,\ps$. 
All systematic uncertainties are added in quadrature, giving a total uncertainty between 4\% and 20\%  of the expected signal yield, depending on the signal mass and lifetime.
In general, the systematic uncertainties have a very limited impact on the measurement. 
The precision on the upper limits is dominated by the statistical uncertainties of the observed yields.


\begin{figure}[tb]
  \begin{center}
    \includegraphics[width=0.85\linewidth]{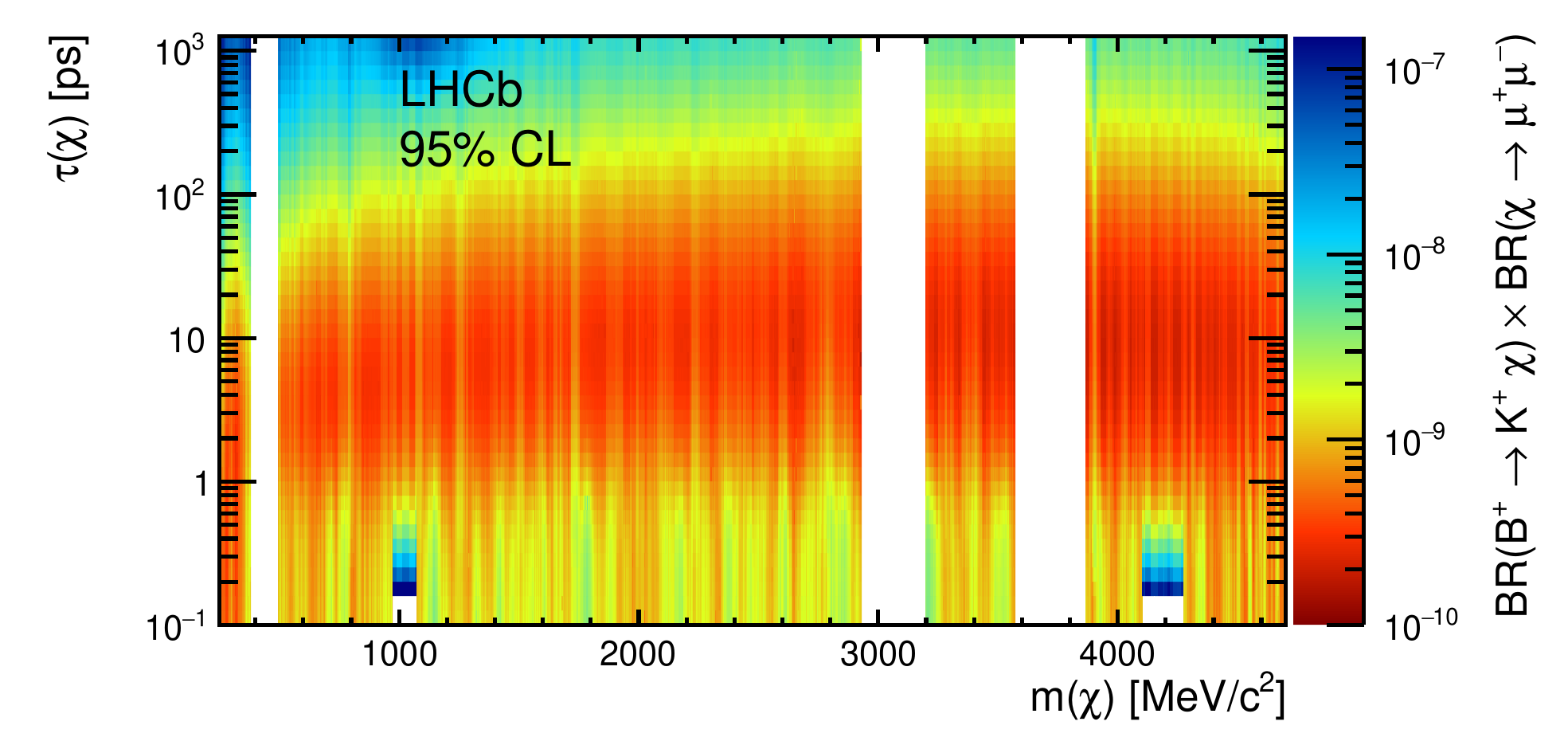}
    \vspace*{-0.7cm}
  \end{center}
  \caption{Excluded branching fraction for the $\Bp \to \Kp \chi (\mumu)$ decay as a function of $m(\chi)$ and $\tau(\chi)$ at 95\% CL. Regions corresponding to the fully-vetoed \KS, \jpsi, \psitwos and \psiprpr and to the partially-vetoed \phiz and $\psi(4160)$ are excluded from the figure. All systematic uncertainties are included in the calculation of the upper limit.}
  \label{fig:2Dlimit}
\end{figure}

Figure~\ref{fig:2Dlimit} shows the upper limits, at 95\% confidence level (CL), on the branching fraction for the decay $\Bp \to \Kp \chi (\mumu)$ as a function of $m(\chi)$ and $\tau(\chi)$.
The upper limits vary between $2\times10^{-10}$ and $10^{-7}$ and are most stringent in the region around $\tau(\chi) = 10\,\ps$.
For longer lifetimes the limit becomes weaker as the probability for the $\chi$ to decay within the vertex detector decreases.
Nevertheless, the present analysis improves previous limits by up to a factor of 20 in the region of long lifetimes $\tau(\chi) \sim 1000\,\ps$. The main improvement with respect to the previous search~\cite{LHCB-PAPER-2015-036} comes from the optimisation of the three regions of the dimuon decay time and the fact that the previous analysis studied the mode $\Bz \to \Kstarz \chi$ , which has in general a smaller branching fraction than the decay $\Bp \to \Kp \chi$, and where only the decays of $\Kstarz \to \Kp \pim$ were reconstructed.


\begin{figure}[tb]
  \begin{center}
    \includegraphics[width=0.67\linewidth]{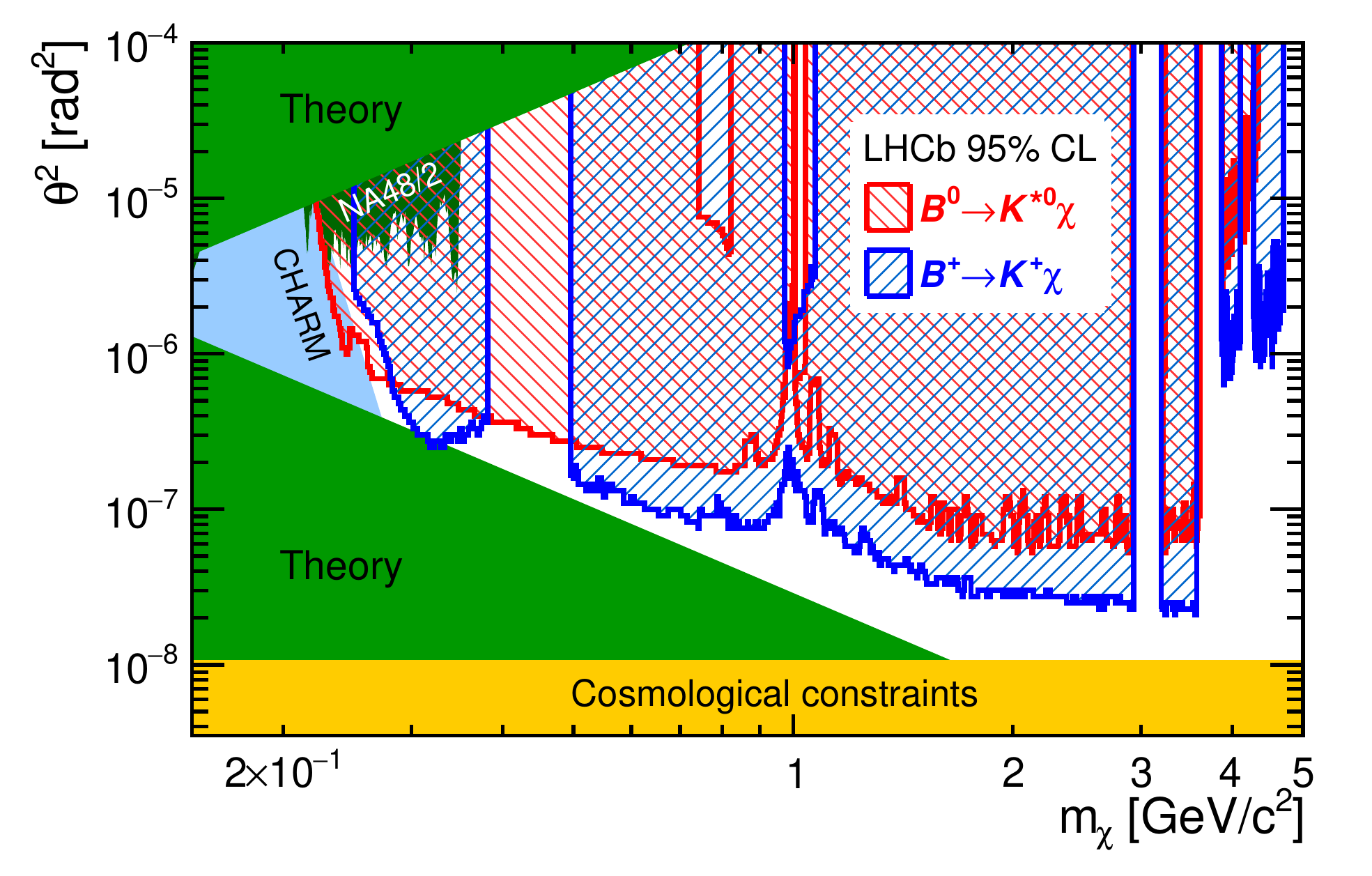}
    \vspace*{-0.7cm}
  \end{center}
  \caption{Parameter space of the inflaton model described in Refs.~\cite{Batell:2009jf,inflatonGuide,Bezrukov:2013fca}. 
  The region excluded at 95\% CL by this analysis is shown by the blue hatched area.
  The region excluded by the search with the $\Bz \to \Kstarz \chi (\mumu)$ decay~\cite{LHCB-PAPER-2015-036} is indicated by the red hatched area.
  Direct experimental constraints set by the CHARM experiment~\cite{CHARM} and NA48 experiments~\cite{CHARM,CERNNA48/2:2016tdo} and regions forbidden by theory or cosmological constraints~\cite{Bezrukov:2013fca} are also shown. 
  }
  \label{fig:limit_inflaton}
\end{figure}

Figure~\ref{fig:limit_inflaton} shows the excluded region at 95\% CL of the parameter space of the inflaton model presented in Refs.~\cite{Batell:2009jf,inflatonGuide,Bezrukov:2013fca}.  
Constraints are placed on the square of the mixing angle, $\theta^2$, which appears in the inflaton effective coupling to the SM fields via mixing with the Higgs boson. 
The inflaton lifetime is predicted to scale as  $\tau \propto 1/\theta^2$.
The $\Bp \to \Kp \chi$ branching fraction is taken from Ref.~\cite{Batell:2009jf}. It is predicted to be between $10^{-4}$ and $10^{-8}$ in the explored region and scales as $\BF(\Bp \to \Kp \chi) \propto \theta^2$,
while the inflaton branching fraction into muons is directly taken from Figure~3 of Ref.~\cite{Bezrukov:2013fca} and is predicted to be between 100\% and slightly less than 1\%, depending on the kinematically allowed decay channels.
Figure~\ref{fig:limit_inflaton} also presents the theoretical and cosmological constraints~\cite{Bezrukov:1GeVinfl,Bezrukov:2013fca} and previous limits set by the CHARM~\cite{CHARM} and the \lhcb~\cite{LHCB-PAPER-2015-036} experiments. 

In summary, a search for a long-lived scalar particle has been performed at \lhcb  using $pp$ collision data corresponding to an integrated luminosity of $3\,\invfb$. 
No evidence for a signal over the background-only hypothesis has been found and upper limits have been placed on $\mathcal{B}(\PB^+ \to \PK^+ \chi) \times \mathcal{B}(\chi \to \mu^+ \mu^-)$. 
They are the best upper limits on this decay to date, improving previous limits by up to a factor of 20. 
The results imply stringent constraints on theories that predict the existence of new light scalar particles. 
For the case of the inflaton model studied here, a large fraction of the theoretically allowed parameter space has been excluded.

\section*{Acknowledgements}

\noindent We express our gratitude to our colleagues in the CERN
accelerator departments for the excellent performance of the LHC. We
thank the technical and administrative staff at the LHCb
institutes. We acknowledge support from CERN and from the national
agencies: CAPES, CNPq, FAPERJ and FINEP (Brazil); NSFC (China);
CNRS/IN2P3 (France); BMBF, DFG and MPG (Germany); INFN (Italy);
FOM and NWO (The Netherlands); MNiSW and NCN (Poland); MEN/IFA (Romania);
MinES and FASO (Russia); MinECo (Spain); SNSF and SER (Switzerland);
NASU (Ukraine); STFC (United Kingdom); NSF (USA).
We acknowledge the computing resources that are provided by CERN, IN2P3 (France), KIT and DESY (Germany), INFN (Italy), SURF (The Netherlands), PIC (Spain), GridPP (United Kingdom), RRCKI and Yandex LLC (Russia), CSCS (Switzerland), IFIN-HH (Romania), CBPF (Brazil), PL-GRID (Poland) and OSC (USA). We are indebted to the communities behind the multiple open
source software packages on which we depend.
Individual groups or members have received support from AvH Foundation (Germany),
EPLANET, Marie Sk\l{}odowska-Curie Actions and ERC (European Union),
Conseil G\'{e}n\'{e}ral de Haute-Savoie, Labex ENIGMASS and OCEVU,
R\'{e}gion Auvergne (France), RFBR and Yandex LLC (Russia), GVA, XuntaGal and GENCAT (Spain), Herchel Smith Fund, The Royal Society, Royal Commission for the Exhibition of 1851 and the Leverhulme Trust (United Kingdom).


\addcontentsline{toc}{section}{References}
\setboolean{inbibliography}{true}
\bibliographystyle{LHCb}

\ifx\mcitethebibliography\mciteundefinedmacro
\PackageError{LHCb.bst}{mciteplus.sty has not been loaded}
{This bibstyle requires the use of the mciteplus package.}\fi
\providecommand{\href}[2]{#2}

\newpage
\centerline{\large\bf LHCb collaboration}
\begin{flushleft}
\small
R.~Aaij$^{40}$,
B.~Adeva$^{39}$,
M.~Adinolfi$^{48}$,
Z.~Ajaltouni$^{5}$,
S.~Akar$^{59}$,
J.~Albrecht$^{10}$,
F.~Alessio$^{40}$,
M.~Alexander$^{53}$,
S.~Ali$^{43}$,
G.~Alkhazov$^{31}$
% ...

{\footnotesize \it
% ...
$ ^{5}$Clermont Universit{\'e}, Universit{\'e} Blaise Pascal, CNRS/IN2P3, LPC, Clermont-Ferrand, France\\
% ...
$ ^{10}$Fakult{\"a}t Physik, Technische Universit{\"a}t Dortmund, Dortmund, Germany\\
% ...
$ ^{31}$Petersburg Nuclear Physics Institute (PNPI), Gatchina, Russia\\
% ...
$ ^{39}$Universidad de Santiago de Compostela, Santiago de Compostela, Spain\\
$ ^{40}$European Organization for Nuclear Research (CERN), Geneva, Switzerland\\
% ...
$ ^{43}$Nikhef National Institute for Subatomic Physics, Amsterdam, The Netherlands\\
% ...
$ ^{48}$H.H. Wills Physics Laboratory, University of Bristol, Bristol, United Kingdom\\
% ...
$ ^{53}$School of Physics and Astronomy, University of Glasgow, Glasgow, United Kingdom\\
% ...
$ ^{59}$University of Cincinnati, Cincinnati, OH, United States\\
% ...
}
\end{flushleft}




\end{document}